\documentclass[twocolumn,secnumarabic,amssymb, nobibnotes, aps, prl]{revtex4-1}

\usepackage{epsfig}
\usepackage{graphicx}
\usepackage{dcolumn}
\usepackage{bm}

\begin{document}

\title{Orbital-dependent electron dynamics in Fe-pnictide superconductors}

\author{Ganesh Adhikary$^1$}
\altaffiliation{Corresponding authors: ganesh.adhikary@ung.si, giovanni.de.ninno@ung.si, kbmaiti@tifr.res.in}
\author{Barbara Ressel$^1$}
\author{Matija Stupar$^1$}
\author{Primo$\check{z}$ Rebernik Ribi$\check{c}$$^2$}
\author{Jurij Urban$\check{c}$i$\check{c}$$^1$}
\author{Giovanni De Ninno$^1$$^,$$^2$$^\ast$}
\author{D. Krizmancic$^3$}
\author{A. Thamizhavel$^4$}
\author{Kalobaran Maiti$^4$$^\ast$}
\affiliation{$^1$Laboratory of Quantum Optics, University of Nova Gorica, 5001 Nova Gorica, Slovenia.}
\affiliation{$^2$Elettra-Sincrotrone Trieste, Area Science Park, 34149 Trieste, Italy.}
\affiliation{$^3$Laboratorio TASC, IOM-CNR, SS 14 Km 163.5, Basovizza, I-34149 Trieste, Italy.}
\affiliation{$^4$Department of Condensed Matter Physics and Materials' Science, Tata Institute of Fundamental Research, Homi Bhabha Road, Colaba, Mumbai - 400 005, INDIA.}

\begin{abstract}
We report on orbital-dependent quasiparticle dynamics in EuFe$_{2}$As$_{2}$, a parent compound of Fe-based superconductors, and a novel way to experimentally identify this behavior using time- and angle-resolved  photoelectron spectroscopy across the spin density wave transition. We observe two different relaxation time scales for photo-excited $d_{xz}/d_{yz}$ and $d_{xy}$ electrons. While the itinerant $d_{xz}/d_{yz}$ electrons relax faster through the electron-electron scattering channel, $d_{xy}$ electrons form a quasi-equilibrium state with the lattice due to their localized character, and the state decays slowly. Our findings suggest that electron correlation in Fe-pnictides is an important property, which should be taken into careful account when describing the electronic properties of both parent and carrier-doped compounds, and therefore establish a strong connection with cuprates.
\end{abstract}

\date{November 20, 2017}


\maketitle

Electron correlation plays a key role in the mechanism of high-temperature superconductivity in cuprates \cite{correlation1,correlation2}. Strong electron correlation is indeed responsible for both Mott behavior of parent compounds and high-$T_{c}$ superconductivity upon electron or hole doping \cite{cuprate}. Since the discovery of high-$T_{c}$ superconductivity in Fe-pnictides  \cite{Kamihara1, Kamihara2}, extensive research has been carried out to understand the pairing mechanism in these systems \cite{Iron-Pnictide}. While several studies point to the existence of strongly correlated electrons in the presence of hole doping \cite{kfe2as2}, electron correlation has hardly been detected in parent and electron-doped compounds  \cite{zxshen_arpes, ga_eufeas, bafe2as2_lda}.

One of the complicating factors in Fe-based compounds is that multiple bands (or orbitals) form the Fermi surface, which makes these systems more complex than cuprates, where the contribution to the Fermi surface comes only from the $d_{x^{2}-y^{2}}$ orbital \cite{cuprate, Iron-Pnictide}.  In Ref. \cite{de_medici} - De Medici {\it et al.} have proposed orbital selective Mottness as an explanation of the unconventional properties of Fe-based superconductors. Each orbital shows single-band Mott behavior, where the degree of electron correlation depends on the doping of the bands from half-filling. Such orbital decoupling and differentiation of correlation strength among different orbitals is caused by Hund's rule that prevents inter-orbital coupling. The theory is also supported by recent results obtained on LiFeAs \cite{orbital_correlation} and several Fe-chalcogenide systems using angle-resolved photo-emission spectroscopy (ARPES), where it was shown that 3$d_{xy}$ electrons are localized, whereas those associated with other 3$d$ orbitals are itinerant \cite{zxshen} in nature. Additional evidence for orbital-selective Mott phase behavior in Fe-chalcogenide systems comes from THz spectroscopy \cite{thz_spectroscopy}, Hall measurements \cite{hall}, pump-probe spectroscopy \cite{pp}, and high-pressure transport measurements \cite{transport}. Theoretical calculations within the multiorbital Hubbard model on K$_{1-x}$Fe$_{2-y}$Se$_{2}$ also supports this picture \cite{kfeas}.

However, for Fe-pnictides, different experiments and theory  show that the differentiation of the correlation strength among orbitals is low for the parent compound and decreases further with electron doping \cite{de_medici, electron_doped_bafe2as2}. Since superconductivity emerges even in electron-doped compounds where all the electrons appear to be itinerant, this raises the question whether strong electron correlation is really the key ingredient required to explain the unconventional properties of the Fe-pnictide family.

Recently, several time-resolved pump-probe experiments were carried out in order to study optically excited states in Fe-pnictides \cite{pump_probe1, pump_probe2, bovensiepen_prl} and gain further understanding of the electronic properties of these systems. For instance, pump-probe data on superconducting Ba$_{1-x}$K$_{x}$Fe$_{2}$As$_{2}$ samples \cite{pump_probe2} revealed a fast and a slow relaxation time-scale for photoexcited carriers, which were associated, respectively, to recombination of quasiparticles through interband and intraband processes. These studies were not able to disentangle the role played by different orbitals, but confirmed that the electronic properties of these systems stem from the multiband nature of the Fermi surface.

In this Letter, we propose an orbital-selective method allowing to study the relaxation dynamics of photo-excited electrons near the Fermi level, and use it as a powerful tool to gain a deep insight into electron localization in Fe-pnictides. In essence, by using time- and angle-resolved photoelectron spectroscopy (trARPES) on EuFe$_{2}$As$_{2}$, a parent compound of the Fe-pnictide family, we observe two different relaxation time scales for electrons in $d_{xz}/d_{yz}$ and $d_{xy}$ orbitals, selectively probed by adjusting the polarization of the pump laser \cite{cuprate}. The slow (fast) relaxation dynamics of excited $d_{xy}$ ($d_{xz}/d_{yz}$) electrons are associated with their localized (itinerant) nature. Our results demonstrate that the differentiation of the correlation strength among orbitals in Fe-pnictides might be higher than previously thought.

The electronic structure of Fe-based superconductors typically consists of three hole bands at the Brillouin zone center ($\Gamma$ point), and two electron bands at the Brillouin zone corner ($M$ point); the number of distinct hole/electron bands can vary depending on the degeneracy of the involved electronic eigenstates. The calculated electronic band structure of EuFe$_2$As$_2$ is shown in Fig. 1. The energy bands shown in Fig. 1(a) exhibit three hole pockets around $\Gamma$-point denoted by $\alpha$, $\beta$ and $\gamma$. The electron pockets around $M$-point are denoted by $\xi$ and $\epsilon$. The inner electron pocket denoted by $\epsilon$ at $M$-point, shows a good nesting condition with the $\beta$ band \cite{Richard-nesting}. It is well established that the nesting condition between the electron pocket and the hole pockets formed by the $\beta$ band causes the spin density wave (SDW) transition in these compounds \cite{khadiza-PRB18}. In Fig. 1(b), we show the schematic of the surface Brillouin zone for (001) plane along with the nesting vector, $Q_{nesting}$. The schematic exhibiting the opening of the SDW gap below the N\'{e}el temperature, 190 K due to the nesting of the $\beta$ and $\epsilon$ bands is shown in Fig. 1(c). The partial density of states of $d_{xy}$ and degenerate ($d_{xz}$/$d_{yz}$) states are shown in Fig. 1(d); the width of the $d_{xy}$ band is significantly narrower than the width of the other bands exhibiting relatively stronger electron correlation induced effects for the $d_{xy}$ electrons.

\begin{figure}
\includegraphics[scale=0.5]{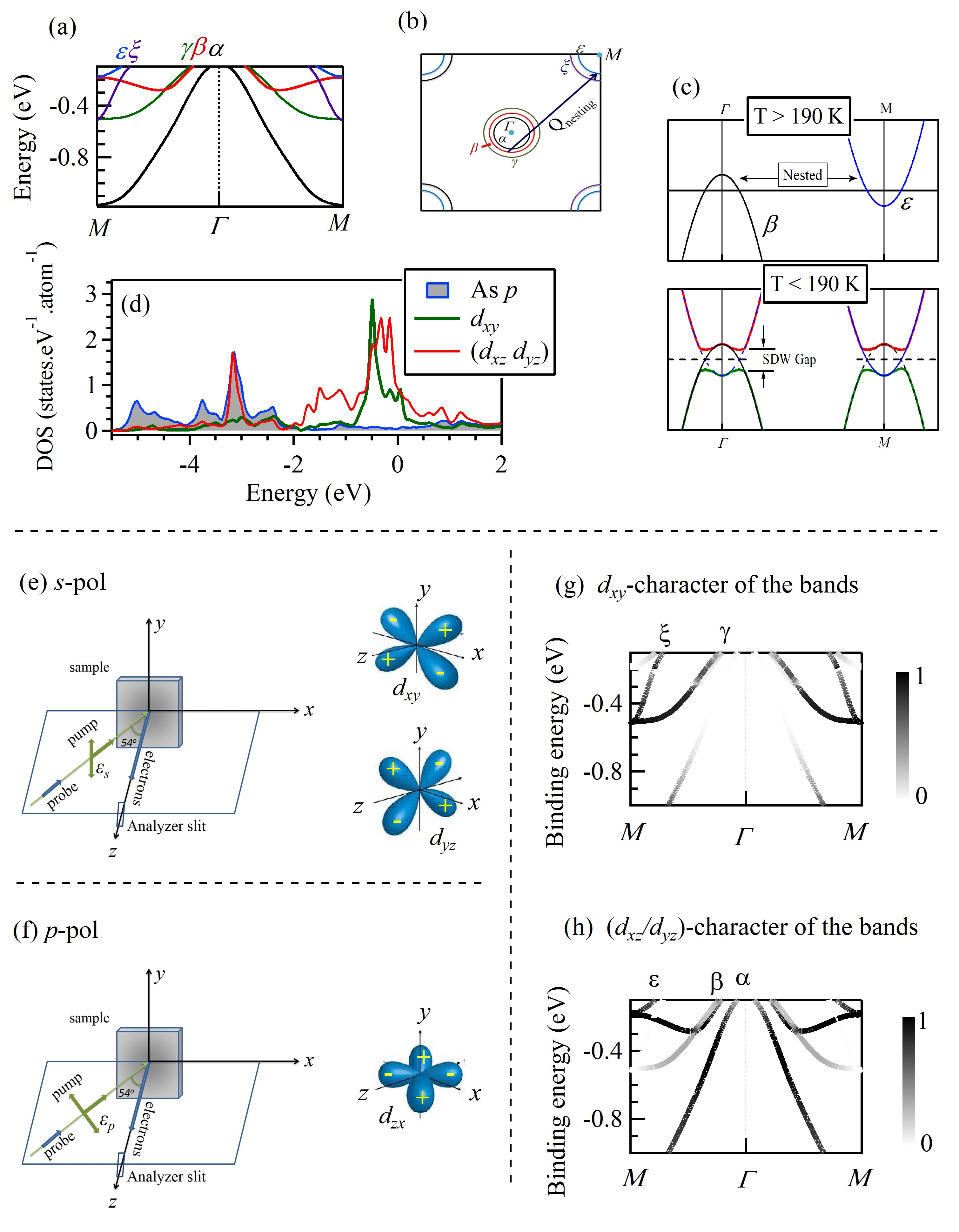}
 \vspace{-2ex}
\caption{(color online) (a) Calculated energy bands of EuFe$_2$As$_2$ along $M\Gamma M$ direction exhibiting $\alpha$, $\beta$ and $\gamma$ bands forming the hole pockets and $\epsilon$, $\xi$ bands forming the electron pockets. (b) Schematic of the surface Brillouin zone and the nesting vector, $Q_{nesting}$ for the SDW phase. (c) Schematic showing the opening of the SDW gap below 190 K due to the nesting of $\beta$ and $\epsilon$ bands. (d) Partial density of states of $d_{xy}$ and degenerate $d_{xz}/d_{yz}$ states. Schematic of the experimental geometry for the excitation by (e) $s$-polarized pump pulse along with the orientation of $d_{xy}$ and $d_{yz}$ orbitals, and (f) $p$-polarized pump pulse along with the orientation of the $d_{xz}$ orbital. Calculated energy bands with (g) $d_{xy}$ character and (h) $d_{xz}/d_{yz}$ character.}
 \vspace{-2ex}
\end{figure}

Angle resolved photoemission spectroscopy (ARPES) is a tool of choice for probing the electronic structure of a system. In a time-resolved ARPES (trARPES) experiment, the dynamics of photoexcited electrons can be tracked by varying the delay, $\Delta t$ between a low-energy (visible or infrared) pump and a high-energy (extreme ultraviolet) probe beam. This gives information about different coupling phenomena, excitation modes and relaxation processes. As we demonstrate here, by adjusting the polarization of the excitation pulse, electronic states can be probed selectively. In Figs. 1(e) and 1(f), we show the experimental geometry for the two configurations when the pump is either $s$-polarized ($s$-pol) or $p$-polarized ($p$-pol). The electric dipole moment vector in the $s$-pol case is in-plane of the sample surface, whereas the dipole moment vector for the $p$-pol beam makes an angle of 36$^\circ$ with the sample surface normal. In the $s$-pol configuration, the pump beam excites the electrons having $d_{xy}$ and $d_{yz}$ symmetry as the electric dipole moment vector has a finite component along the orbital lobe of both these orbitals shown in Fig. 1(g) and 1(h), respectively. On the other hand, the $p$-pol configuration of the pump beam will excite predominantly $d_{xz}$ states \cite{orbital_character,fink,polarization} shown in Fig. 1(h).

Time- and angle-resolved photoemission spectroscopy was carried out using a mode-locked Ti:sapphire laser system delivering pulses at 1.5 eV (800 nm), with 50 $fs$ duration and 5 kHz repetition rate. The pulse was split into two parts. The major part of the intensity was used to produce high order harmonics, spanning the energy range from 10 to 50 eV, using argon as the generating medium \cite{citius}. The second part of the beam was used as a pump, whose intensity was controlled with a variable attenuator based on a half wave-plate and a polarizer. We could select the desired harmonics and control their flux by means of a specially-designed grating set-up, which preserves the pulse duration. The probe energy was set to 29 eV. At this energy, the photoionization cross-section of Fe 3$d$ states is higher compared to As 4$p$ states \cite{cross_section}. The use of a relatively high photon energy also enables probing larger $k$-range in the reciprocal space for a fixed acceptance angle of the electron analyzer. At lower photon energies, the photoionization cross-section of As 4$p$ states is much higher compared to Fe 3$d$ states and the contribution from secondary electrons becomes significant. The photoemission chamber was equipped with an R3000 analyzer from VG Scienta, 5-axis manipulator and a closed cycle He-cryostat from Prevac. Single crystalline EuFe$_2$As$_2$ samples were grown by the Sn-flux method \cite{growth_procedure}. The samples were cleaved {\it in-situ} at a pressure of about 5$\times$10$^{-10}$ Torr to generate a clean and flat surface before each measurement. The measurements were done at 1$\times$10$^{-10}$ Torr. The probe polarization was fixed to $s$-pol and the pump polarization was adjusted using a polarizer.

\begin{figure}
\includegraphics[scale=0.5]{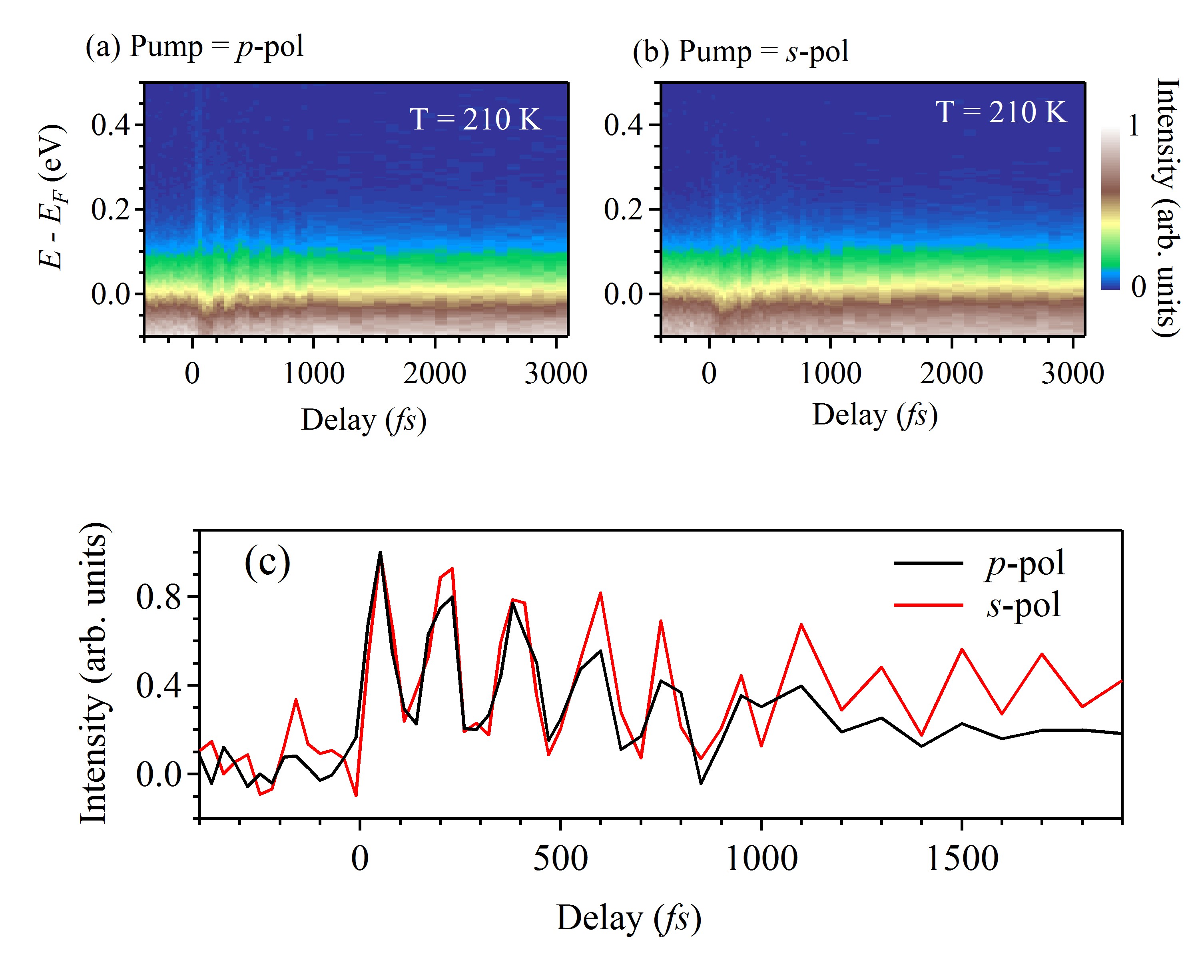}
 \vspace{-2ex}
\caption{(color online) Time resolved photoemission spectra as a function of $\Delta t$ at 210 K using (a) $p$-polarized ($p$-pol) and (b) $s$-polarized ($s$-pol) pump pulses. (c) Intensity of the hot electrons (integrated between 0.1 - 0.2 eV above the Fermi level) as a function of $\Delta t$ for both pump polarizations. Significant difference in decay is observed for $\Delta t > 1 ps$.}
 \vspace{-2ex}
\end{figure}

In Figs. 2(a) and 2(b), we show the time-resolved photoemission spectra as a function of  the delay, $\Delta t$ between the pump and the probe pulses at 210 K ($T > T_N$) for both the polarizations, $p$-pol and $s$-pol, respectively. The intensity of hot electrons was obtained integrating intensities within 0.1 to 0.2 eV binding energy above the Fermi level level in order to avoid contributions from thermally excited electrons within the Fermi-Dirac distribution and/or energy resolution broadening. The results show sharp rise in intensity at $\Delta t$ = 0 (corresponding to temporal overlap of the pump and probe pulses) followed by coherent oscillations, and a decay of the signal in the $ps$ time-scale (see Fig. 2(c)). After optical excitation of electrons to unoccupied states, electron-electron scattering and electron-lattice interactions give rise to different coherent collective excitation modes such as phonons and magnons within tens of femtosecond. The frequency of oscillations is 5.6 THz (around 23 meV) for both polarizations, which corresponds to the fully symmetric $A_{1g}$ phonon mode triggered by the breathing of As atoms along $c$-axis \cite{jpcm_bovensiepen}. It is worth stressing that we have not observed significant change in the electron dynamics with pump fluence within the range of 2 mJ/cm$^{2}$ - 5 mJ/cm$^{2}$, as these pulse energies are far too low to excite the system into the anharmonic regime \cite{phonon}. While coherent oscillations seem to be similar for both polarizations at lower $\Delta t$, we observe significant difference in the electron relaxation dynamics for $\Delta t >~1~ps$; the intensity of the signal decreases faster in $p$-pol configuration compared to the $s$-pol case. Furthermore, the data in the $p$-pol configuration show faster damping of the oscillations.

\begin{figure}
\includegraphics[scale=0.4]{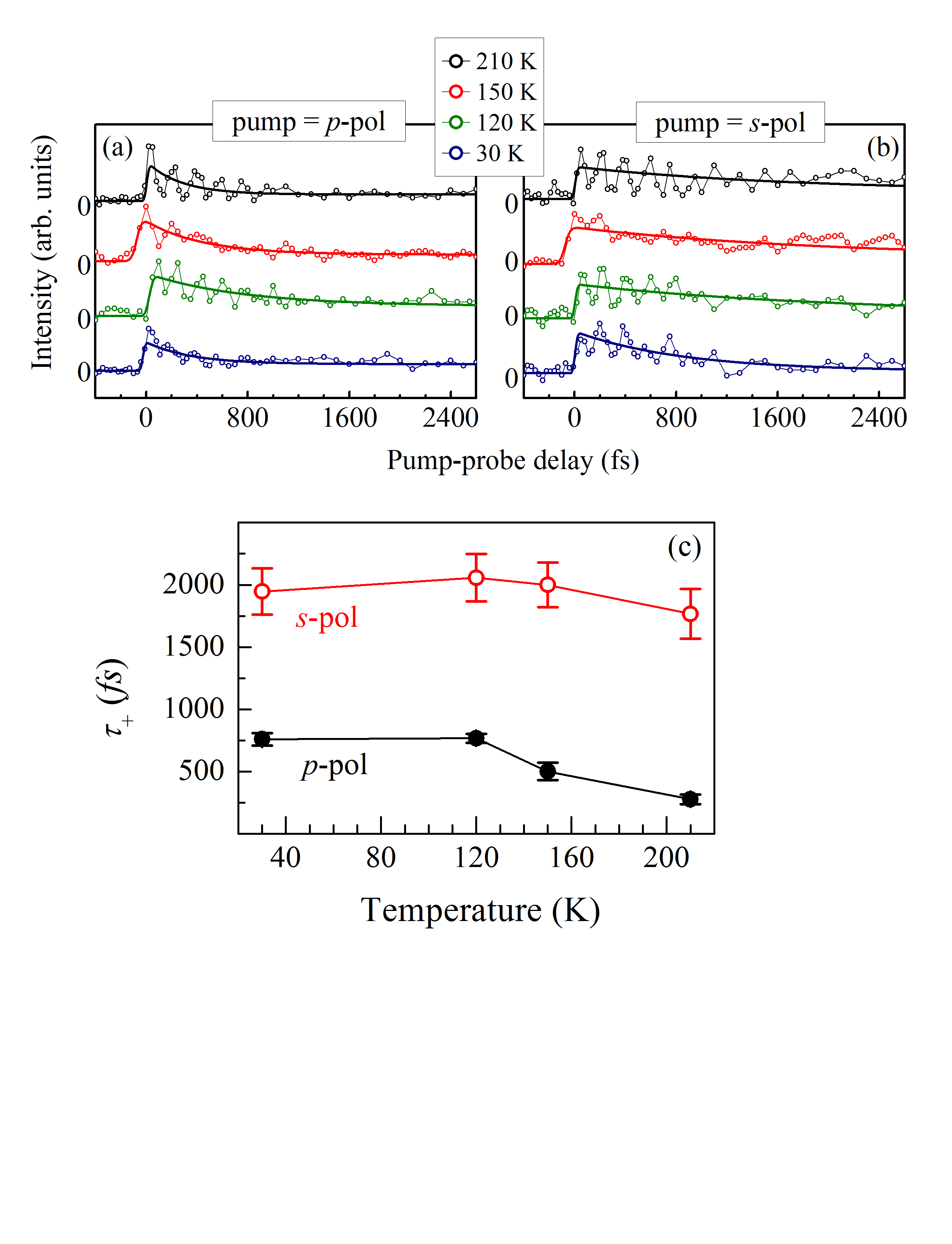}
 \vspace{-18ex}
\caption{(color online) Spectral intensity of the hot electrons as $\Delta t$ for (a) $p$-polarized and (b) $s$-polarized pump pulses at different sample temperatures. Solid lines represent single exponential fits. (c) Decay time constants of the hot electrons, $\tau_{+}$ as a function of temperature for $p$-pol and $s$-pol pump pulse. Significantly different time constants for $p$-pol and $s$-pol responses demonstrate successful polarization selection of different decay channels by the polarized pump pulse.}
 \vspace{-2ex}
\end{figure}

The intensity of the hot electrons, $I_+$ above and below $T_N$ is shown in Fig. 3 for both $p$-pol and $s$-pol pump excitations. To analyze the temporal dynamics of the hot holes ($I_{-}$), we integrated the signal in a 100 meV energy window below the Fermi level. The time-dependent intensity profiles of the hot holes for $p$- and $s$-polarizations of the pump at different temperatures are shown in Fig. 4. We used $I_{+,-} = A exp(-t/\tau_{+,-}) + B$ as a fit function allowing to extract the decay time constant of the excited states. Here, $A$ is the amplitude of the excitation, $\tau_{+,-}$ is the decay time constant of the excited electrons or holes, respectively and $B$ accounts for the background originating from electron-phonon scattering. The fitting function was multiplied by a step function at $\Delta t = 0$ and convoluted with a Gaussian function to account for finite durations of the pump and probe pulses. The extracted decay constants vs. temperature for hot electrons and holes are shown in Figs. 3(c) and 4(c), respectively for both polarizations. The hot electron dynamics are significantly different for the two pump polarizations as shown in Fig. 3(c). The decay is fast with a time constant of approximately 500 $fs$, when pumped by $p$-pol light. For the $s$-pol pump pulse, the decay is slow with a time constant in the range of 1-2 $ps$. Furthermore, the dynamics of hot electrons shows considerably (a factor of 3) slower relaxation in the SDW phase compared to high temperatures, when pumped by $p$-pol, while for $s$-pol pumping, changes with temperature are not significant.

\begin{figure}
\includegraphics[scale=0.4]{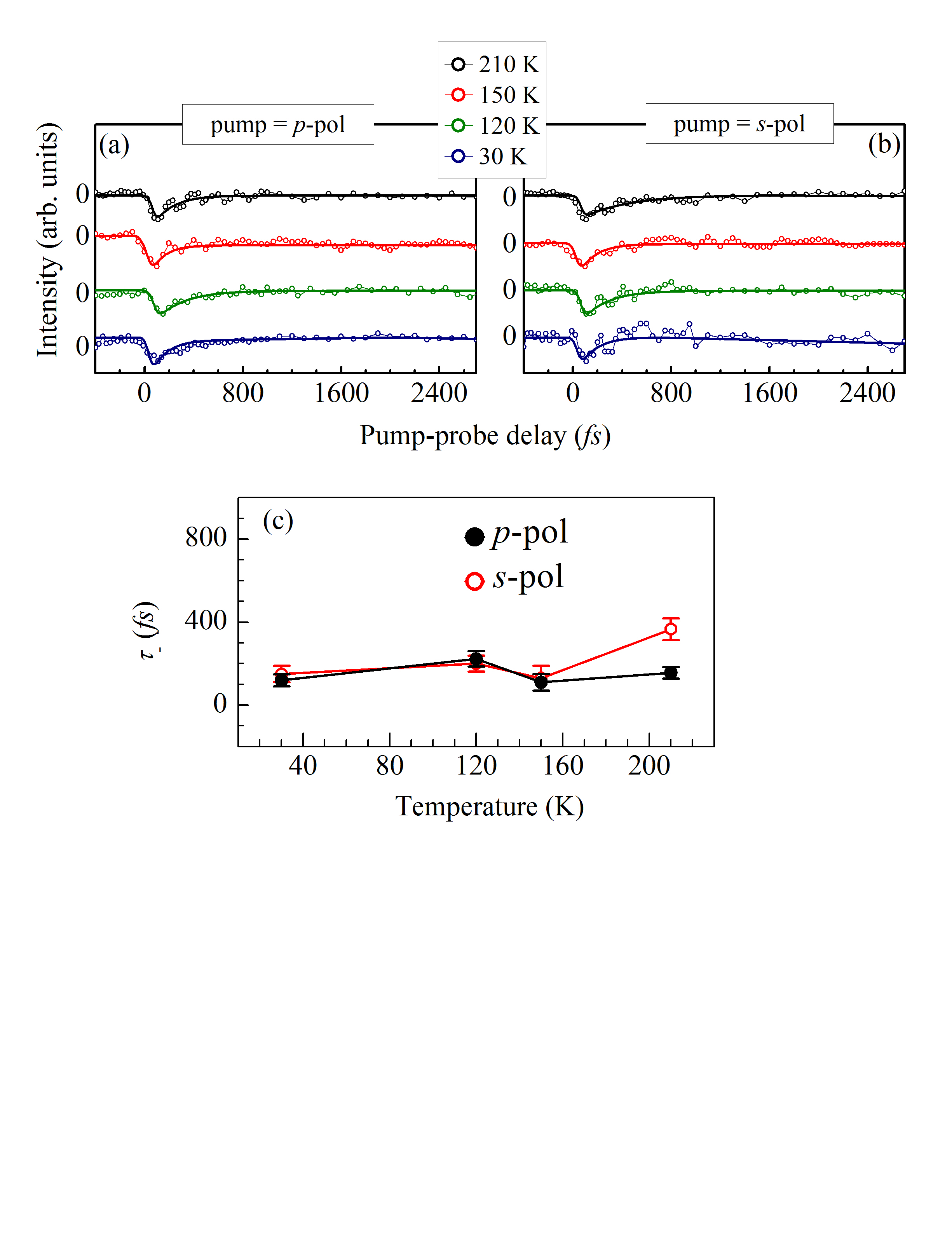}
 \vspace{-28ex}
\caption{(color online) Spectral intensity of hot holes as a function of $\Delta t$ for (a) $p$-polarized and (b) $s$-polarized pump pulses at different sample temperatures. Solid lines represent single exponential fits. (c) Hole decay time constants, $\tau_{-}$ for $p$-polarized and $s$-polarized pump pulses as a function of temperature. While the hole for $s$-pol case survives longer than the hole for $p$-pol case at 210 K, this difference in decay times becomes negligible in the SDW phase.}
 \vspace{-2ex}
\end{figure}

In the $p$-pol configuration, the pump beam primarily excites electrons in $d_{xz}$ orbitals as demonstrated in Fig. 1(h). Their fast relaxation dynamics can be explained by strong interband scattering between the $\beta$ band and the electron pocket at the $M$-point ($\epsilon$ band) \cite{pump_probe2}, which is possible due to their good Fermi surface nesting. We argue that such fast relaxation dynamics, i.e., the ability to efficiently dissipate energy, can be associated to the itinerant (delocalized) nature of $d_{xz}$ electrons.  On the other hand, the relaxation dynamics observed for $s$-polarized light is complex. At short delay time, it seem to follow a trend akin to the $p$-pol case, but the difference becomes significant at longer delay; hot electrons survive for a much longer time in the $s$-pol case. As demonstrated in Fig. 1(e), the $s$-polarized pump pulse can excite both $d_{yz}$ and $d_{xy}$ states. Thus, the data in the shorter delay time seem to have influence from the decay of $d_{yz}$ states, while the longer delay time is predominantly contributed by the decay of $d_{xy}$ states, which is not present in the $p$-pol case. As the $\gamma$ band has no nesting condition with any other band, $d_{xy}$ electrons can only decay through intraband scattering and hence are longer lived.

A similar scenario also manifests in the hole dynamics by the difference in decay time in the paramagnetic phase (210 K, see Fig. 4(c)), where the behavior of the $d_{xy}$ and $d_{xz}/d_{yz}$ orbitals are significantly different. The decay of holes is primarily governed by the energy transfer to the lattice through electron-phonon relaxation. This is manifested by somewhat faster relaxation of hot holes with decreasing temperature as more and more phonon modes are available for energy transfer at lower temperatures and thus enhances the phase space for electron-phonon scattering. The decay of $d_{xz}$ holes are found to be less sensitive to temperature due to various competing effects such as opening of SDW gap, nematicity (lifting the degeneracy of $d_{xz}d_{yz}$ bands), change in structural parameters, etc.

We can obtain further insight into the electron relaxation dynamics by resorting to the Rothwarf-Taylor model \cite{rtmodel}. According to the model, in a compound with an energy gap of 2$\Delta$, the decay rate of the excited states depends on a number of processes. First, the recombination of the quasi-particles (QPs) across the SDW gap can take place. This process creates a photon of energy 2$\Delta$. In the second step, the emitted photon can recreate another QP-pair, create a low energy boson (such as a phonon) or escape out of the probed region. In the normal (non-SDW phase), only the latter two processes can take place. On the other hand, when the system is brought into the SDW state, a gap opens up at the Fermi level in the $\beta$ band. This generates a relaxation bottleneck due to QP recombination and recreation, i.e., the QPs may form a quasi-equilibrium state with the lattice, which we observe as an increase of the decay time constant at lower temperatures (Fig. 3(c)) in the case of $p$-polarized pump. In the $s$-pol configuration, where the decay is primarily governed by the relaxation of $d_{xy}$ electrons, we observe no significant change in the decay time constant as the $\gamma$ band does not participate in the SDW transition.

In summary, we have demonstrated a powerful method, complementary to the measurement of electron effective mass, allowing to probe orbital-selective Mott phase behavior in strongly correlated materials. Specifically, we observe two different time-scales for the relaxation of electrons in $d_{xz}$/$d_{yz}$ and $d_{xy}$ orbitals in EuFe$_{2}$As$_{2}$, a parent compound of Fe-based superconductors. We find that $d_{xz}$/$d_{yz}$ electrons relax fast through electron-electron scattering. Such fast relaxation dynamics can be attributed to the itinerant nature of these orbitals. On the other hand, $d_{xy}$ electrons are found to relax over significantly longer time scales. We associate the slow dynamics to the fact that such electrons create a quasi-equilibrium state with the lattice due to their high degree of localization. Although the ratio between effective masses of electrons belonging to $d_{xz}$/$d_{yz}$ and $d_{xy}$ measured with different techniques, such as ARPES and quantum oscillations, is $\sim$~1 \cite{mass, mass1, mass2}, our results show that the nature of electrons in these two types of orbitals is quite different. Our findings suggest that orbital dependent electron correlation in Fe-pnictide is important and should be taken into careful account when describing the electronic properties of both parent and carrier doped compounds, and therefore establish a strong connection with the cuprates.

KM acknowledges financial assistance from the Department of Science and Technology, Government of India under the J.C. Bose Fellowship program and the Department of Atomic Energy under the DAE-SRC-OI Award.


\begin{thebibliography}{99}
%
\bibitem{correlation1} A. Garg, M. Randeria, And N. Trivedi, Nat. Phys. {\bf 4}, 762 (2008).
%
\bibitem{correlation2} C. Weber, K. Haule, and G. Kotliar, Nat. Phys. {\bf 6}, 574 (2010).
%
\bibitem{cuprate} A. Damascelli, Z. Hussain, and Z.-X. Shen, Rev. Mod. Phys. {\bf 75}, 473 (2003).
%
\bibitem{Kamihara1} Y. Kamihara, H. Hiramatsu, M. Hirano, R. Kawamura, H. Yanagi, T. Kamiya, and H. Hosono, J. Am. Chem. Soc. {\bf 128}, 10012 (2006).
%
\bibitem{Kamihara2} Y. Kamihara, T. Watanabe, M. Hirano, and H. Hosono, J. Am. Chem. Soc. {\bf 130}, 3296 (2008).
%
\bibitem{Iron-Pnictide} K. Ishida, Y. Nakai, and H. Hosono, J. Phys. Soc. Jpn. {\bf 78}, 062001 (2009).
%
\bibitem{kfe2as2} F. Hardy, A. E. B$\ddot{o}$hmer, D. Aoki, P. Burger, T. Wolf, P. Schweiss, R. Heid, P. Adelmann, Y. X. Yao, G. Kotliar, J. Schmalian, and C. Meingast, Phys. Rev. Lett. {\bf 111}, 027002 (2013); G. Adhikary, D. Biswas, N. Sahadev, S. Ram,
V. Kanchana, C. S. Yadav, P. L. Paulose, and K. Maiti, J. Appl. Phys. {\bf 114}, 163906 (2013).
%
\bibitem{zxshen_arpes} M. Yi, D. H. Lu, J. G. Analytis, J.-H. Chu, S.-K. Mo, R.-H. He, M. Hashimoto, R. G. Moore, I. I. Mazin, D. J. Singh, Z. Hussain, I. R. Fisher, and Z.-X. Shen, Phys. Rev. B {\bf 80}, 174510 (2009).
%
\bibitem{bafe2as2_lda} M. Yi, D. H. Lu, J. G. Analytis, J.-H. Chu, S.-K. Mo, R.-H. He, R. G. Moore, X. J. Zhou, G. F. Chen, J. L. Luo, N. L. Wang, Z. Hussain, D. J. Singh, I. R. Fisher, and Z.-X. Shen, Phys. Rev. B {\bf 80}, 024515 (2009).
%
\bibitem{ga_eufeas} G. Adhikary {\it et al.}, J. Phys.: Condens. Matter {\bf 25}, 225701 (2013); G. Adhikary {\it et al.}, J. Appl. Phys. {\bf 115}, 123901 (2014).
%
\bibitem{de_medici} L. d\'{e} Medici, G. Giovannetti, and M. Capone, Phys. Rev. Lett. {\bf 112}, 177001 (2014).
%
\bibitem{orbital_correlation} T. Hajiri, T. Ito, M. Matsunami, B. H. Min, Y. S. Kwon, K. Kuroki, and S. Kimura, Phys. Rev. B {\bf 93}, 024503 (2016).
%
\bibitem{zxshen} M. Yi, Z-K Liu, Y. Zhang, R. Yu, J.-X. Zhu, J. J. Lee, R. G. Moore, F. T. Schmitt, W. Li, S. C. Riggs, J.-H. Chu, B. Lv, J. Hu, M. Hashimoto, S.-K. Mo, Z. Hussain, Z. Q. Mao, C. W. Chu, I. R. Fisher, Q. Si, Z.-X. Shen, and D. H. Lu, Nat. Commun. {\bf 6}, 7777 (2015).
%
\bibitem{thz_spectroscopy} Z. Wang, M. Schmidt, J. Fischer, V. Tsurkan, M. Greger, D. Vollhardt, A. Loidl, and J. Deisenhofer, Nat. Commun. {\bf 5}, 3202 (2014).
%
\bibitem{hall} X. Ding, Y. Pan, H. Yang, and H.-H. Wen, Phys. Rev. B {\bf 89}, 224515 (2014).
%
\bibitem{pp} W. Li, C. Zhang, S. Liu, X. Ding, X. Wu, X. Wang, H.-H. Wen, and M. Xiao, Phys. Rev. B {\bf 89}, 134515 (2014).
%
\bibitem{transport} P. Gao, R. Yu, L. Sun, H. Wang, Z. Wang, Q. Wu, M. Fang, G. Chen, J. Guo, C. Zhang, D. Gu, H. Tian, J. Li, J. Liu, Y. Li, X. Li, S. Jiang, K. Yang, A. Li, Q. Si, and Z. Zhao, Phys. Rev. B {\bf 89}, 094514 (2014).
%
\bibitem{kfeas} R. Yu, P. Goswami, Q. Si, P. Nikolic, and J.-X. Zhu, Nat. Commun {\bf 4}, 2783 (2013).
%
\bibitem{electron_doped_bafe2as2} B. Mansart, E. Papalazarou, M. F. Jensen, V. Brouet, L. Petaccia, L. de' Medici, G. Sangiovanni, F. Rullier-Albenque, A. Forget, D. Colson, and M. Marsi, Phys. Rev. B {\bf 85}, 144508 (2012).
%
\bibitem{pump_probe1} E. E. M. Chia, D. Talbayev, J.-X. Zhu, H. Q. Yuan, T. Park, J. D. Thompson, C. Panagopoulos, G. F. Chen, J. L. Luo, N. L. Wang, and A. J. Taylor, Phys. Rev. Lett. {\bf 104}, 027003 (2010).
%
\bibitem{pump_probe2} D. H. Torchinsky, G. F. Chen, J. L. Luo, N. L. Wang, and N. Gedik, Phys. Rev. Lett. {\bf 105}, 027005 (2010).
%
\bibitem{bovensiepen_prl} L. Rettig, R. Cort$\acute{e}$s, S. Thirupathaiah, P. Gegenwart, H. S. Jeevan, M. Wolf, J. Fink, and U. Bovensiepen, Phys. Rev. Lett. {\bf 108}, 097002 (2012).
%
\bibitem{Richard-nesting} P. Richard {\it et al.}, Phys. Rev. Lett. {\bf 102}, 047003 (2009).
%
\bibitem{khadiza-PRB18}
Khadiza Ali {\it et al.}, Phys. Rev. B {\bf 97}, 054505 (2018); K. Ali and K. Maiti, Scientific Reports {\bf 7}, 6298 (2017).
%
\bibitem{orbital_character} Y. Zhang, F. Chen, C. He, B. Zhou, B. P. Xie, C. Fang, W. F. Tsai, X. H. Chen, H. Hayashi, J. Jiang, H. Iwasawa, K. Shimada, H. Namatame, M. Taniguchi, J. P. Hu, and D. L. Feng, Phys. Rev. B {\bf 83}, 054510 (2011).
%
\bibitem{fink} J. Fink, S. Thirupathaiah, R. Ovsyannikov, H. A. D$\ddot{u}$rr, R. Follath, Y. Huang, S. de Jong, M. S. Golden, Y.-Z. Zhang, H. O. Jeschke, R. Valentí, C. Felser, S. D. Farahani, M. Rotter, and D. Johrendt, Phys. Rev. B {\bf 79}, 155118 (2009).
%
\bibitem{polarization} A. C. Liu, J. St\"{o}hr, C. M. Friend, and R. J. Madix, Surface Science, {\bf 235}, 107 (1990); T. Hajiri, T. Ito, R. Niwa, M. Matsunami, B. H. Min, Y. S. Kwon, and S. Kimura Phys. Rev. B {\bf 85}, 094509 (2012).
%
%
\bibitem{citius} C. Grazioli, C. Callegari, A. Ciavardini, M. Coreno, F. Frassetto, D. Gauthier, D. Golob, R. Ivanov, A. Kivim$\ddot{a}$ki, B. Mahieu, B. Bu$\check{c}$ar, M. Merhar, P. Miotti, L. Poletto, E. Polo, B. Ressel, C. Spezzani, and G. De Ninno, Rev. Sci. Instrum. {\bf 85}, 023104 (2014).
%
\bibitem{cross_section} J. J. Yeh and I. Lindau, At. Data and Nucl. Data Tables {\bf 32}, 1 (1985).
%
\bibitem{growth_procedure} Y. Xiao, Y. Su, M. Meven, R. Mittal, C. M. N. Kumar, T. Chatterji, S. Price, J. Persson, N. Kumar, S. K. Dhar, A. Thamizhavel, and Th. Brueckel, Phys. Rev. B {\bf 80}, 174424 (2009).
%
\bibitem{jpcm_bovensiepen} I. Avigo, R. Cor$\acute{t}$es, L. Rettig, S. Thirupathaiah, H. S. Jeevan, P. Gegenwart, T. Wolf, M. Ligges, M. Wolf, J. Fink, and U. Bovensiepen, J. Phys.: Condens. Matter {\bf 25} 094003 (2013).
%
\bibitem{phonon} B. Mansart, D. Boschetto, A. Savoia, F. Rullier-Albenque, A. Forget, D. Colson, A. Rousse, and M. Marsi, Phys. Rev. B {\bf 80}, 172504 (2009).
%
\bibitem{rtmodel} A. Rothwarf and B. N. Taylor, Phys. Rev. Lett. {\bf 19}, 27 (1967).
%
\bibitem{mass} S. L. Skornyakov, A. V. Efremov, N. A. Skorikov, M. A. Korotin, Yu. A. Izyumov, V. I. Anisimov, A. V. Kozhevnikov, and D. Vollhardt, Phys. Rev. B {\bf 80}, 092501 (2009).
%
\bibitem{mass1} J. G. Analytis, R. D. McDonald, J.-H. Chu, S. C. Riggs, A. F. Bangura, C. Kucharczyk, M. Johannes, and I. R. Fisher, Phys. Rev. B {\bf 80}, 064507 (2009).
%
\bibitem{mass2} S. E. Sebastian, J. Gillett, N Harrison, P. H. C. Lau, D. J. Singh, C. H. Mielke, and G. G. Lonzarich, J. Phys.: Condens. Matter {\bf 20}, 422203 (2008).
%


\end{thebibliography}
\end{document}